\documentclass{article}
\usepackage{ismir,amsmath,cite}
\usepackage{aliascnt}
\usepackage{tabu}
\usepackage{amssymb,amsmath,mathtools}
\usepackage{booktabs}
\usepackage{multirow}

\usepackage{times}
\usepackage{url}
\usepackage{color}
\usepackage{multirow}
\usepackage{multicol}
\usepackage{graphicx}
\usepackage{caption}
\usepackage{tikz}
\usetikzlibrary{calc,tikzmark}
\usepackage[shortcuts]{extdash}
\usepackage{blindtext}
\usepackage[ruled,norelsize,linesnumbered]{algorithm2e}

\makeatletter
\def\inmod#1{\allowbreak\mkern5mu{\operator@font mod}\,\,#1}
\makeatother

\title{A predictive model for music based on learned interval representations}
\multauthor
{Stefan Lattner$^{1,2}$, Maarten Grachten$^{1,2}$, Gerhard Widmer$^1$}
{$^1$ Institute of Computational Perception, JKU Linz\\
$^2$ Sony Computer Science Laboratories (CSL), Paris, France}
\def\authorname{Stefan Lattner, Maarten Grachten, Gerhard Widmer}
\begin{document}

\maketitle

\begin{abstract}
Connectionist sequence models (e.g., RNNs) applied to musical sequences suffer from two known problems:
First, they have strictly ``absolute pitch perception''.
Therefore, they fail to generalize over musical concepts which are commonly perceived in terms of \emph{relative distances} between pitches (e.g., melodies, scale types, modes, cadences, or chord types).
Second, they fall short of capturing the concepts of repetition and musical form.
In this paper we introduce the \emph{recurrent gated autoencoder} (RGAE), a recurrent neural network which learns and operates on \emph{interval representations} of musical sequences.
The relative pitch modeling increases generalization and reduces sparsity in the input data.
Furthermore, it can learn sequences of copy-and-shift operations (i.e. chromatically transposed copies of musical fragments)---a promising capability for learning musical repetition structure.
We show that the RGAE improves the state of the art for general connectionist sequence models in learning to predict monophonic melodies, and that ensembles of relative and absolute music processing models improve the results appreciably.
Furthermore, we show that the relative pitch processing of the RGAE naturally facilitates the learning and the generation of sequences of copy-and-shift operations, wherefore the RGAE greatly outperforms a common absolute pitch recurrent neural network on this task.

\end{abstract}


\section{Introduction}\label{sec:introduction}
The objective of sequence models for music prediction is to predict (the probability of) musical events at the next time step, given some prior musical context.
In the (most common) case of predicting note events, this task involves finding relationships between past and future occurrences of absolute pitches.
However, many music theoretical constructs that might help to find such relationships are defined in relative terms, such as diatonic scale steps, and cadences.
The discrepancy between the relative nature of many regularities in music and the absolute pitch representation is problematic for modeling tasks, because it leads to high sparsity in the input data, increased model sizes, and altogether reduced generalization in music modeling.

To remedy these problems, musical input sequences can be transposed to a common key before training, augmented by random transpositions during training, or, in case of symbolic monophonic music, transformed into interval representations before training.
In this work, we propose a sequence model which \emph{learns} both interval representations from absolute pitch sequences and temporal dependencies between these intervals.
By learning not only the intervals between two successive notes, but all intervals within a window of $n$ pitches, the model is more robust to diatonic transposition and can also learn repetition structure.
More precisely, a recurrent neural network (RNN) is employed on top of a gated autoencoder (GAE), which we refer to as \emph{recurrent gated autoencoder} (RGAE).
The GAE portion learns the intervals between its input and its target pitches and represents them in its latent space.
The RNN portion operates on these interval representations, to learn their temporal dependencies.
The implicit transformation to intervals allows this architecture to operate directly on absolute musical textures, without the need for data preprocessing.
Besides, relative pitch modeling reduces the sparsity in the data and the representations learned by the GAE are transposition-invariant.
Therefore, the RGAE requires less temporal connections than a common RNN while achieving higher prediction accuracy.

Also, operating on the intervals of input sequences brings added value to sequence modeling.
By allowing the model to relate its prediction with events using specific time lags, it can learn copy-and-shift operations.
In the space of intervals, such operations are performed by repeatedly applying a constant interval to events occurring a constant time lag in the past.
Moreover, the RNN portion of the architecture can learn sequences of such copy-and-shift operations (i.e., ``structure schemes''), which can then be realized as musical notes by the GAE.

This ability is promising for music modeling, where musical form defines the self-similarity within a piece, and repeated sections often occur as a transposed (i.e., shifted in the pitch dimension) version of the initial section.
Musical form is challenging to learn with common sequence models, like RNNs.
They are specialized in learning the statistics of musical textures and are ``blind'' towards similarity and (transposed) repetition (i.e., there is no content-independent ``repetition neuron'').
As a result, when sampling music using such models, repeated fragments occur either due to chance or as a phenomenon of an entanglement with a learned texture.
In contrast, the ability of RGAEs to learn copy-and-shift operations may allow to represent musical form explicitly, and to realize learned schemes as musical textures in music prediction and music generation tasks.



We show that the RGAE is competitive with state-of-the-art models in a music sequence learning task.
Furthermore, we demonstrate that the RGAE, due to its relative pitch processing, is complementary to absolute pitch models, by combining their predictions to obtain improved accuracy.
Lastly, we show that the RGAE is particularly suited for learning sequences of copy-and-shift operations.
It can learn to recognize and continue pre-defined ``structure schemes'', abstracted from the actual texture, with \linebreak which the scheme is realized.

In Section \ref{sec:related_work}, we provide an overview of related models and related publications.
In Section \ref{sec:models}, the GAE and the proposed extensions to the RGAE are described, as well as the baseline RNN used for comparison and combined prediction.
General training details concerning the GAE are given in Section \ref{sec:training}.
The two experiments conducted, including the data used, training details and discussion for each experiment separately, are presented in Section \ref{sec:experiments}.
Section \ref{sec:conclusion} concludes the paper and provides further directions.

\vspace{-3mm}
\section{Related Work}\label{sec:related_work}
GAEs are bi-linear models utilizing \emph{multiplicative interactions} to learn correlations between or within data instances.
They were introduced by \cite{memisevic2011gradient} as a derivative of the gated Boltzmann machines (GBMs) \cite{memisevic2007unsupervised, memisevic2010learning}, as standard learning criteria became applicable through the development of denoising autoencoders \cite{vincent2010stacked}.
In music, bi-linear models were applied to learn co-variances within spectrogram data for music similarity estimation \cite{schluter2011music}, and for learning musical transformations in the symbolic domain \cite{lattner2017relations}.

The GAE was utilized for learning the derivatives of sequences in \cite{memisevic2013aperture} (between subsequent frames in movies of rotated 3D objects), and to predict accelerated motion by stacking two layers to learn second-order derivatives \cite{michalski2014modeling}.
This method is very similar to the one proposed here, but we use different dimensionalities between input and output, and we do not assume constant transformations but rather learn \emph{sequences of transformations} using an RNN.

Probabilistic n-gram models, specialized on learning to predict monophonic pitch sequences include IDyOM \cite{pearce2005construction}, and \cite{DBLP:conf/ismir/LanghabelLTR17}, both employ multiple features of the musical surface.
In this paper, we do not compare the RGAE with these models, as they are more specialized on the musical domain, by explicit selection of (computed) features.
We compare the RGAE to the currently best performing general connectionist sequence model, the RTDRBM \cite{cherla2016neural}.
Its architecture is similar to the well-known RTRBM proposed in \cite{DBLP:conf/nips/SutskeverHT08}, but it employs a different cost function.

For structured sequence generation, Markov chains together with pre-defined repetition structure schemes were employed in \cite{collins2016developing}, where specific methods for handling transitions between repeating segments were proposed; in \cite{pachet2017sampling}, where an approach to a controlled creation of variations was introduced; in \cite{conklin_semiotic}, where chords were generated, obeying a pre-defined repetition structure.
In \cite{lattnergeneration}, a convolutional restricted Boltzmann machine was employed, and different structural properties were imposed using differentiable soft-constraints and gradient descent optimization.
A constrained variable neighborhood search to generate polyphonic music obeying a tension profile and the repetition structure from a template piece was proposed in \cite{herremans2016morpheus}.
In \cite{eigenfeldt2013evolving}, Markov chains and evolutionary algorithms were used to generate repetition structure for Electronic Dance Music.

\vspace{-3mm}
\section{Models}\label{sec:models}
\subsection{Gated Autoencoder}\label{sec:gae-training}
A GAE learns first-order derivatives between its input and its output.
In musical sequences, this amounts to learning pitch intervals, which are represented as distinct codes in its latent space.
In reconstruction, it applies learned interval codes to pitches in order to transpose them.
Its ability to learn and to perform musical transformations is, however, not limited to single intervals.
For example, it was shown in \cite{lattner2017relations}, that more complex musical transformations like diatonic transposition can be learned by a GAE and can be applied to an unseen material.
Intervals are encoded in the latent space of the GAE, denoted as mappings
\begin{equation}\label{eq:gamap}
\mathbf{m}_{t+1} = \sigma_q(\mathbf{W}_m (\mathbf{Q}\mathbf{x}_{t-n}^{t} \cdot \mathbf{V}\mathbf{x}_{t+1})),
\end{equation}
where $\mathbf{x}_{t+1}$ is a binary vector encoding active notes at time step $t+1$ as on-bits, $\mathbf{x}_{t-n}^{t}$ contain the concatenated vectors of the last $n$ time steps, $\mathbf{Q}, \mathbf{V}$ and $\mathbf{W}_m$ are weight matrices, and $\sigma_q$ is the softplus non-linearity.
The operator~$\cdot$ (indicated as a triangle in Figure \ref{fig:rgae}) depicts the Hadamard product of the filter responses $\mathbf{Q}\mathbf{x}_{t-n}^{t}$ and $\mathbf{V}\mathbf{x}_{t+1}$, denoted as \emph{factors}.
This operation allows the model to \emph{relate} its inputs, making it possible to learn interval representations.

GAEs are often trained by minimizing the symmetric error when reconstructing the output from the input and vice versa.
In the proposed RGAE architecture, we use predictive training and just learn to reconstruct the target $\mathbf{x}_{t+1}$ from the input $\mathbf{x}_{t-n}^{t}$ and the mapping $\mathbf{m}_{t+1}$ as
\begin{equation}\label{recony}
\mathbf{\tilde{x}}_{t+1} = \sigma_g(\mathbf{V}^\top (\mathbf{W}_m^\top \mathbf{m}_{t+1} \cdot \mathbf{Q}\mathbf{x}_{t-n}^{t})),
\end{equation}
where $\sigma_g$ is the sigmoid non-linearity.
The GAE portion of the RGAE is pre-trained by minimizing the binary cross-entropy loss of the reconstruction as
\begin{equation}\label{eq:cost}
\mathcal{L}(\mathbf{x},\mathbf{\tilde{x}}) = -\frac1N\sum_{n=1}^N\ \bigg[x_n  \log_2 \tilde x_n + (1 - x_n)  \log_2 (1 - \tilde x_n)\bigg]\,.
\end{equation}

\subsection{Recurrent Gated Autoencoder}\label{sec:model}
\begin{figure}
\begin{center}
\includegraphics[width=\linewidth]{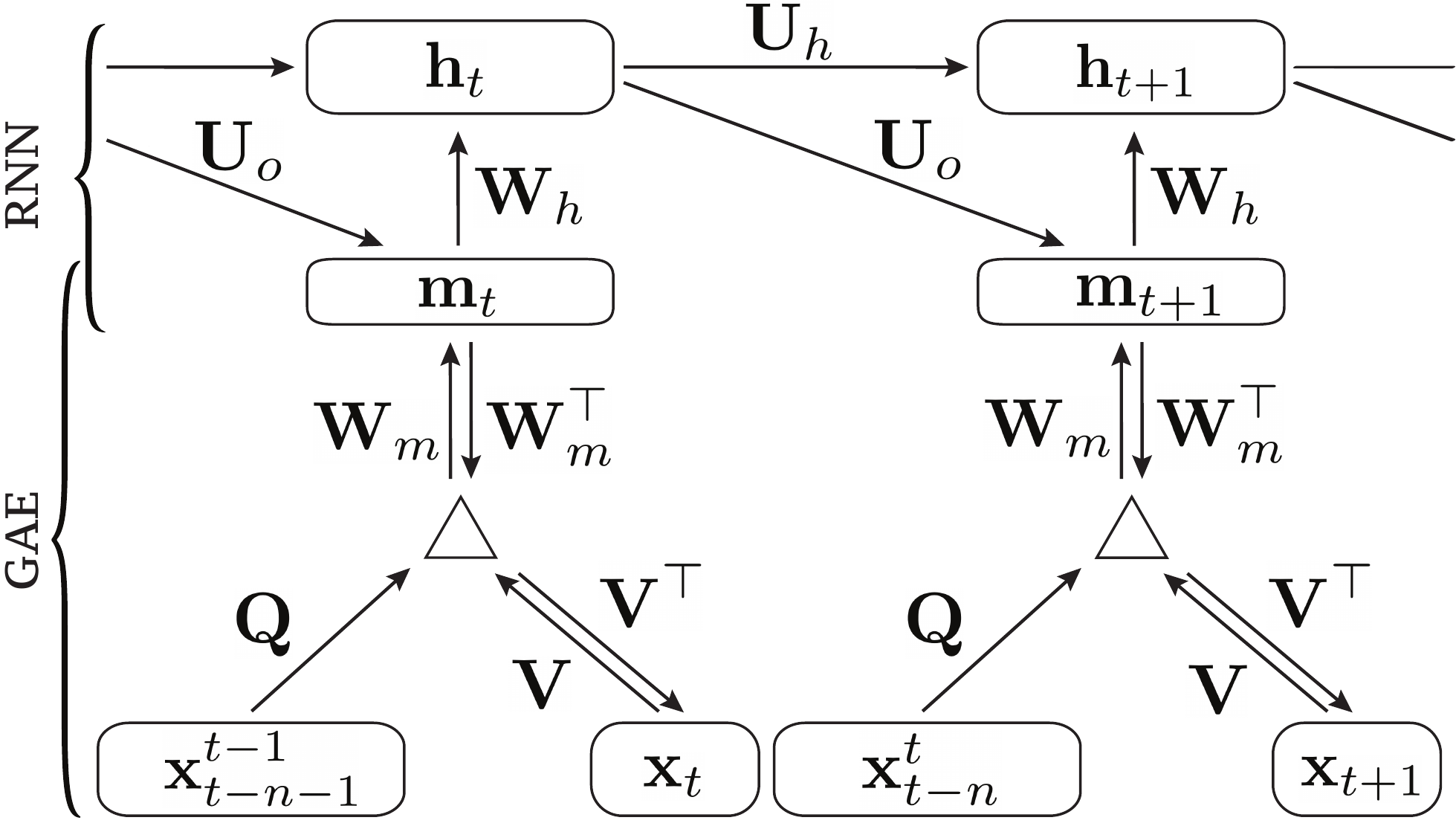}
\caption{Schematic illustration of the proposed recurrent gated autoencoder architecture. Arrows represent weight matrices, rounded rectangles represent vectors. The triangles depict the Hadamard product. The specifics of the gated recurrent unit are omitted for better clarity.}
\label{fig:rgae}
\end{center}
\vspace{-3mm}
\end{figure}

The proposed model is a combination of a gated autoencoder (GAE) and a recurrent neural network (RNN) as depicted in Figure \ref{fig:rgae}.
The GAE learns relative pitch (i.e., interval) representations of the musical surface, and the RNN learns their temporal dependencies.

We use gated recurrent units (GRUs) \cite{cho2014properties} for the RNN portion of the RGAE. This type of units have been shown to be often as efficient as long short-term memory units (LSTMs, \cite{hochreiter1997long}) while being conceptually simpler \cite{chung2014empirical}.
It is intuitively clear that any RNN variant can be potentially attached on a GAE.
The input to the RNN at time t is the GAE's mapping $\mathbf{m}_{t}$, resulting in the following specification:
\begin{equation}\label{eq:z}
\mathbf{z}_t = \sigma_g(\mathbf{W}_{z} \mathbf{m}_{t} + \mathbf{U}_{z} \mathbf{h}_{t-1} + \mathbf{b}_z), \\
\end{equation}
\begin{equation}\label{eq:r}
\mathbf{r}_t = \sigma_g(\mathbf{W}_{r} \mathbf{m}_{t} + \mathbf{U}_{r} \mathbf{h}_{t-1} + \mathbf{b}_r), \\
\end{equation}
\begin{equation}\label{eq:h}
\mathbf{h}_t =  \mathbf{z}_t \cdot \mathbf{h}_{t-1} + (1-\mathbf{z}_t) \cdot \sigma_h(\mathbf{W}_{h} \mathbf{m}_{t} + \mathbf{U}_{h} (\mathbf{r}_t \cdot \mathbf{h}_{t-1}) + \mathbf{b}_h),
\end{equation}
where $\mathbf{h}_t$ is the hidden state at time $t$, $\mathbf{z}_t$ is the update gate vector, $\mathbf{r}_t$ is the reset gate vector, and $\mathbf{W}$, $\mathbf{U}$ and $\mathbf{b}$ are parameter matrices and vectors.
The RNN predicts the next mapping of the GAE as
\begin{equation}\label{eq:pred}
\mathbf{\widetilde{m}}_{t+1} = \sigma_q(\mathbf{U}_o \mathbf{h}_t),
\end{equation}
which is used to reconstruct the target configuration at $t+1$ as
\begin{equation}\label{recony_rnn}
\mathbf{\tilde{x}}_{t+1} = \sigma_s(\mathbf{V}^\top (\mathbf{W}_m^\top \mathbf{\widetilde{m}}_{t+1} \cdot \mathbf{Q}\mathbf{x}_{t-n}^{t}))\,.
\end{equation}

Here, we use the softmax non-linearity $\sigma_s$, as the data the RGAE is trained on is monophonic.
The full architecture is trained with Backpropagation through time (BPTT) to minimize the \emph{categorical} cross-entropy loss for the reconstructed target as
\begin{equation}\label{eq:cost_softmax}
\mathcal{L}(\mathbf{x},\mathbf{\tilde{x}}) = -\frac1N\sum_{n=1}^N\ x_n  \log_2 \tilde x_n\,.
\end{equation}

When the RGAE is applied to polyphonic music, in Equation~\ref{recony_rnn} the sigmoid non-linearity, together with the binary cross-entropy loss (cf. Equation \ref{eq:cost}) has to be used.

\subsection{Baseline RNN}
As a baseline, we employ an RNN with GRUs to directly operate on the data.
Accordingly, Equations \ref{eq:z}, \ref{eq:r}, and \ref{eq:h} are adapted to take $\mathbf{x}_t$ instead of $\mathbf{m}_{t}$ as input.
Consequently, the prediction of the baseline RNN amounts to
\begin{equation}\label{eq:pred_rnn}
\mathbf{\tilde{x}}_{t+1} = \sigma_s(\mathbf{U}_o \mathbf{h}_t),
\end{equation}
where the softmax non-linearity is applied, making the categorical cross-entropy loss (cf. Equation \ref{eq:cost_softmax}) applicable in training.


\section{Gated Autoencoder pre-training}\label{sec:training}


Due to the relatively high number of parameters in its GAE portion, the RGAE is prone to overfitting.
To circumvent this, and to establish robust interval representations, we pre-train the GAE first, using the cross-entropy of the reconstruction as the cost function (cf. Equation \ref{eq:cost}).
In the second training iteration, we train the RNN portion of the GAE to minimize the cross-entropy error of the architecture's prediction (cf. Equation \ref{eq:cost_softmax}).
The datasets may differ between the training iterations as long as the included relations are identical (e.g. ``intervals of western tonal music'').
Consequently, the GAE parameters trained on one dataset can be used for prediction tasks on several datasets.
Fine-tuning the whole architecture in the last few epochs of predictive training can make up for possible bias.

In the following, we describe how the GAE is pre-trained in our experiments.
Details varying between the experiments are given later in the experiments section (cf. Section \ref{sec:experiments}).

\subsubsection{Enforcing Transposition-Invariance}

A property of interval representations in music is transposition invariance (i.e., transposing the melody does not change the representation).
Although training the GAE as described in Section \ref{sec:gae-training} naturally tends to lead to similar mapping codes for input target pairs that have the same interval relationships, the training does not explicitly enforce such similarities and consequently the mappings may not be maximally transposition invariant.
Therefore, when pre-training the GAE, we explicitly support the learning of transposition-invariant codes.
First, we define a transposition function $\textit{shift}(\mathbf{x}, \delta)$, which shifts the bits of a vector $\mathbf{x}$ of length $M$ by $\delta$ pitches:
\begin{equation}\label{eq:shift}
\textit{shift}(\mathbf{x}, \delta) = (x_{(0+\delta)\inmod{M}}, \dots, x_{(M-1+\delta)\inmod{M}})^\top,
\end{equation}
where $\textit{shift}(\mathbf{x}_{t-n}^{t}, \delta)$ denotes the transposition of each single time step vector \emph{before} concatenation and linearization.

The altered training is then as follows:
First, the mapping code $\mathbf{m}_{t+1}$ of an input/target pair is inferred as shown in Equation \ref{eq:gamap}. Then, $\mathbf{m}_{t+1}$ is used to reconstruct a \emph{transposed} version of the target from an equally \emph{transposed} input (modifying Equation \ref{recony}) as
\begin{equation}\label{reconyshift}
\mathbf{\tilde{x}}'_{t+1} = \sigma_g(\mathbf{V}^\top (\mathbf{W}_m^\top \mathbf{m}_{t+1} \cdot \mathbf{Q} \textit{shift}(\mathbf{x}_{t-n}^{t},\delta))),
\end{equation}
with $\delta \in [-30,30]$.
Finally, we penalize the error between the reconstruction of the transposed target and the actual transposed target (i.e., employing Equation \ref{eq:cost}) as
\begin{equation}
\mathcal{L}(\textit{shift}(\mathbf{x}_{t+1},\delta),\mathbf{\tilde{x}}'_{t+1}).
\end{equation}

The transposition distance $\delta$ is randomly chosen for each training batch.
This method amounts to both, a form of guided training and data augmentation.


\subsubsection{Pre-training and Architecture}\label{sec;pretrain-details}
We use $512$ units in the factor layer and $64$ units in the mapping layer of the GAE.
On the latter, sparsity regularization \cite{Lee:2007uz} is applied.
The deviation of the norms of the columns of both weight matrices $\mathbf{U}$ and $\mathbf{V}$ from their average norm is penalized. 
Furthermore, we restrict these norms to a maximum value.
The learning rate is reduced from $0.001$ to $0$ during training, and RMSProp \cite{hinton2012neural} is used.

\section{Experiments}\label{sec:experiments}
\subsection{Experiment 1: Folk Song Prediction}\label{sec:predict_folksongs}
We test the RGAE and RNN in a sequence learning task using the data described in Section \ref{sec:data-efsc}.
In order to make the results comparable, we use the same experiment setup as in \cite{pearce2004improved, cherla2016neural}.

\subsubsection{Data}\label{sec:data-efsc}
The EFSC subset (comprising a total of 54,308 note events) of the Essen Folk Song Collection (EFSC) \cite{TheEssenFolksongC:1995um} constitutes the data for the actual training and evaluation.
It consists of 119 Yugoslavian folk songs, 91 Alsatian folk songs, 93 Swiss folk songs, 104 Austrian folk songs, the German subset \emph{kinder} (213 songs), and 237 songs of the Chinese subset \emph{shanxi}.
The melodies are represented as series of pitches ignoring note durations.

For pre-training the GAE portion of the RGAE, we use a polyphonic Mozart piano music dataset (\cite{widmer2003discovering}, comprising 13 piano sonatas with more than 106,000 notes) in piano-roll representation (i.e., using a regular time grid of 1/8th note resolution, and an active note can span several time steps).
We pre-train on that data because polyphonic music acts as a better regularizer for learning interval representations than monophonic music.

\subsubsection{Training and Architecture}
We use only $16$ hidden units in the RNN portion of the RGAE.
The look-back window of the GAE is $n = 8$ pitches, and we apply $50$\% dropout on the input in pre-training and when training the whole architecture.
We pre-train the GAE for 250 epochs on the Mozart piano pieces (cf. Section \ref{sec:data-efsc}).
Subsequently, the RNN portion is trained for 110 epochs on the interval representations (i.e., mappings provided by the GAE) of the EFSC datasets.
In the last $10$ epochs the whole architecture is fine-tuned.

The baseline RNN with $50$ hidden units is trained for 70 epochs on the EFSC data.
The learning rate scheme is adopted from that described in Section \ref{sec;pretrain-details} for all models.

\begin{table*}[t]
\centering
\footnotesize
\begin{tabular}{lllllllll}
\toprule
    &  RNN & RTDRBM \cite{cherla2016neural} & RGAE & RNN + & RNN + & RTDRBM + \\
Data      & (GRU) & &  & RTDRBM & RGAE & RGAE \\
\midrule
Alsatian folk songs & $2.890$ & $2.897$ &  $2.872$  & $2.844$ & $2.788$ & $\mathbf{2.771}$\\
Yugoslavian folk songs  & $2.717$ & $2.655$ & $2.676$ & $2.617$ & $2.586$ & $\mathbf{2.530}$\\
Swiss folk songs  & $2.954$ & $2.932$ & $2.895$  & $2.851$& $2.831$ & $\mathbf{2.769}$\\
Austrian folk songs  & $3.185$ & $3.259$ &  $3.171$  & $3.163$ & $\mathbf{3.070}$ & $3.085$\ \\
German folk songs & $2.358$ & $2.301$ &  $2.305$  & $2.257$ & $2.233$ & $\mathbf{2.184}$\\
Chinese folk songs & $2.725$ & $2.685$ &  $2.752$  & $2.612$& $2.650$ & $\mathbf{2.595}$\\
\midrule
Average  & $2.805$ & $2.788$ & $2.779$ & $2.724$ & $2.693$ & $\mathbf{2.656}$ \\
\bottomrule
\end{tabular}
\caption{Cross-Entropies of the 10-fold cross validation in the prediction task for different data sets and different models. 
When combining the RGAE with an absolute pitch model (i.e., RNN, RTDRBM), results improve substantially.
The results suggest that absolute and relative pitch models are complementary in the aspects they learn about music and can be effectively used in an ensemble method.}
\label{tab:folksong-results}
\vspace{-.3cm}
\end{table*}

\subsubsection{Combining Model Predictions}\label{sec:geom-mean}
We hypothesize that the RNN and the RGAE are complementary in how they process musical sequences.
For example, the RNN may have better stability in remembering absolute reference pitches, like the tonic of a piece, and is superior in modeling prior probabilities, to keep predictions in a plausible pitch range.
In contrast, the RGAE can make use of structural cues indicating repetitions and can generalize better due to relative pitch processing.
There are several possibilities to combine the predictions of statistical models.
Next to the ad-hoc approach of merely averaging their outputs, we can also use information about the certainty of the models and weight their outputs accordingly.
A measure for the certainty of a prediction is given by the Shannon entropy \cite{shannon2001mathematical}:
\begin{equation}
H(p) = -\sum_{a\in A}{p(a) \log_2 p(a)},
\end{equation}
where $p(a\in \mathcal{A}) = P(\mathcal{X} = a)$ is a probability mass function over a discrete alphabet $\mathcal{A}$.
The method which worked best in our experiments is calculating the entropy-weighted geometric mean of both predictions, as proposed in \cite{pearce2004methods}:
\begin{equation}
p(t) = \frac{1}{R} \prod_{m\in M}{p_m(t)^{w_m}},
\end{equation}
where $p_m(t)$ is the predicted distribution of model $m$ at time $t$, $w_m = H_{\text{relative}}(p_m)^{-b}$ is the weight of model $m$, non-linearly scaled using a bias $b$ (set to $0.5$ in our experiments), and $R$ is a normalization constant.
The relative entropy $H_{\text{relative}}(p_m)$ for model $m$ is given by
\begin{equation}
H_{\text{relative}}(p_m) = \frac{H(p_m)}{H_{\text{max}}(p_m)},
\end{equation}
where $H_{\text{max}}(p_m) > 0$ is the entropy of the probability mass uniformly distributed over the alphabet (indicating maximal uncertainty of the model).

\subsubsection{Evaluation}
Since the datasets are rather small, a fixed training/test set split would lead to a poor estimation of the performance of the models.
Therefore, and in accordance with \cite{pearce2004improved, cherla2016neural}, a 10-fold cross validation is performed for each dataset and the categorical cross-entropy loss (cf. Equation \ref{eq:cost_softmax}) is reported.

\subsubsection{Results and Discussion}
The results are shown in Table \ref{tab:folksong-results}.
The current state-of-the-art results for general connectionist sequence models on the datasets are achieved by the RTDRBM model introduced in \cite{cherla2016neural}.
The results show that the RGAE slightly outperforms the RTDRBM and is clearly superior to the baseline RNN.
Note that the RGAE only has $16$ units for learning temporal dependencies (the GAE portion mainly transforms absolute pitch input to relative pitch representations).
This compactness suggests that the relative processing of music indeed supports generalization by reducing the sparsity in the data.

When combining the predictions of the RGAE with an absolute pitch model (i.e., RNN or RTDRBM) based on the entropy-weighted geometric mean (cf. Section \ref{sec:geom-mean}), a more substantial improvement is achieved than when combining the two absolute pitch models.
This result shows that absolute and relative processing of music are complementary and can, therefore, be effectively used together in an ensemble method.

\subsection{Experiment 2: Copy-and-Shift Operations}
This experiment shall be seen as a proof-of-concept for the RGAEs ability to learning sequences of copy-and-shift operations (i.e., structure schemes).
We oppose our model to an RNN with GRUs, which is known to have difficulties to learn tasks in the form ``whatever has been generated before, now create a (shifted) copy of it''.
The hypothesis is that the RGAE, due to its modeling of intervals, is superior in solving this task.
It has shown in previous studies that it can learn content-invariant transformations between data instances \cite{memisevic2013aperture}, a necessary capability for learning content-invariant structure schemes.

\subsubsection{Data}
In order to obtain a controlled setup for testing the model performances, we construct data obeying different recurring (chromatic) transposition patterns.
To this end, the EFSC dataset is transformed into a piano-roll representation with a resolution of 1/8th note.
From that, short fragments of length $4$, $8$, and $16$ ($\le$ the length of the receptive field of the input to the models) are randomly sampled (rests are omitted).
It is necessary that the RGAE has access to all past events with which the prediction should be related.
Choosing longer fragment lengths than the lengths of the receptive fields yields considerably worse results, also for the baseline RNN, which already performs weakly in this setup.
The fragments are copied and transposed according to some pre-defined transposition schemes (cf. Table \ref{tab:transpatterns}).
For each of the $10$ schemes and fragment lengths, 26 sequences (512 time steps each, resulting in $133\,120$ time steps) are generated, where 20 sequences are used for training, 5 sequences are used for testing and 1 for evaluation.
This results in a total of $600$ sequences for training, $150$ sequences for testing and 30 sequences for evaluation.


\begin{table}[]
\centering
\footnotesize
\begin{tabular}{l}
\toprule
Transposition Schemes \\
\midrule
$\{+5,+5,+5,\dots \}$ \\
$\{+7,+7,+7,\dots \}$ \\
$\{-5,-5,-5,\dots \}$ \\
$\{-7,-7,-7,\dots \}$ \\
$\{+12,-12,+12,\dots \}$ \\
$\{+3,-3,+3,\dots \}$ \\
$\{+4,-4,+4,\dots \}$ \\
$\{+9,-9,+9,\dots \}$ \\
$\{+4,-8,+4,-8,\dots \}$ \\
$\{-4,+8,-4,+8,\dots \}$ \\
\bottomrule
\end{tabular}
\caption{The different relative transposition schemes used in Experiment 2.}
\label{tab:transpatterns}
\end{table}

\subsubsection{Training and Architecture}
The lookback window of the RGAE is $n = 16$ time steps, the RNN portion has 64 units, and we do not use dropout on the input.
For the baseline RNN, we also input the 16 preceding time steps, as this supports copy operations by freeing up memory in the hidden units.
The baseline RNN model size (512 units) is selected by starting from 64 units and always doubling that number until no substantial improvement occurs on the evaluation set.

The GAE portion of the RGAE is pretrained for 50 ep-ochs on the structured sequences described above.
Subsequently, the RGAE is trained for $50$ epochs, holding the parameters of the GAE fixed.
As the data of the pretraining does not differ from the sequences in the prediction task, finetuning is not necessary.

The baseline RNN is trained for 60 epochs.
Again, for both models the learning rate scheme described in Section \ref{sec;pretrain-details} is employed.
Note that in this task, we always randomly transpose the input to the models in all training phases.
Therefore, we need no dropout on the input of the RGAE, and the baseline RNN does not overfit, despite its high number of parameters. 

\subsubsection{Evaluation}
The models have to learn to continue sequences from the test set after exposition to the first $64$ time steps of each sequence.
The experiment is different to typical prediction tasks in that possibly incorrect predictions are fed back to the models, causing errors to accumulate.
To obtain more stable continuations, we do not sample from the predicted distributions of the models, but instead, treat the experiment as a classification task and choose the pitch with the highest predicted probability.
Accordingly, the precision is merely the percentage of correctly predicted pitches over time.
In addition, we quantify how many sequences are correctly continued until the end by considering all sequences with an overall precision above $99\%$ as correctly continued.
Furthermore, like in Experiment 1, the categorical cross-entropy loss (cf. Equation \ref{eq:cost_softmax}) is computed.

\subsubsection{Results and Discussion}\label{sec:discussion}
\begin{figure}
\includegraphics[width=1.\linewidth]{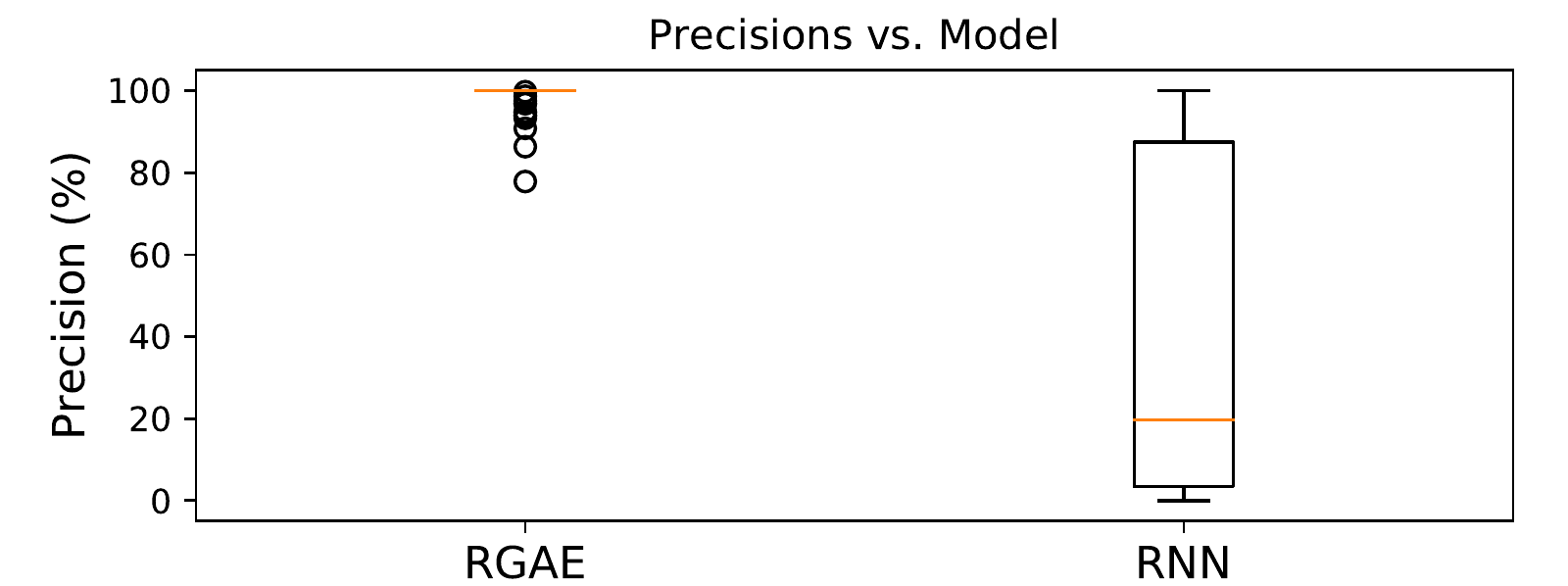} \\
\vspace{-8mm} \\
\caption{Distribution of precisions for continuation of sequential copy-and-shift operations in the test set of size $150$. The median is marked with a orange line, the boxes indicate the interquartile range, and circles indicate outliers.}
\label{fig:prec-boxpl}
\vspace{-.3cm}
\end{figure}

\begin{figure}[t]
\begin{center}
\includegraphics[width=1.\linewidth]{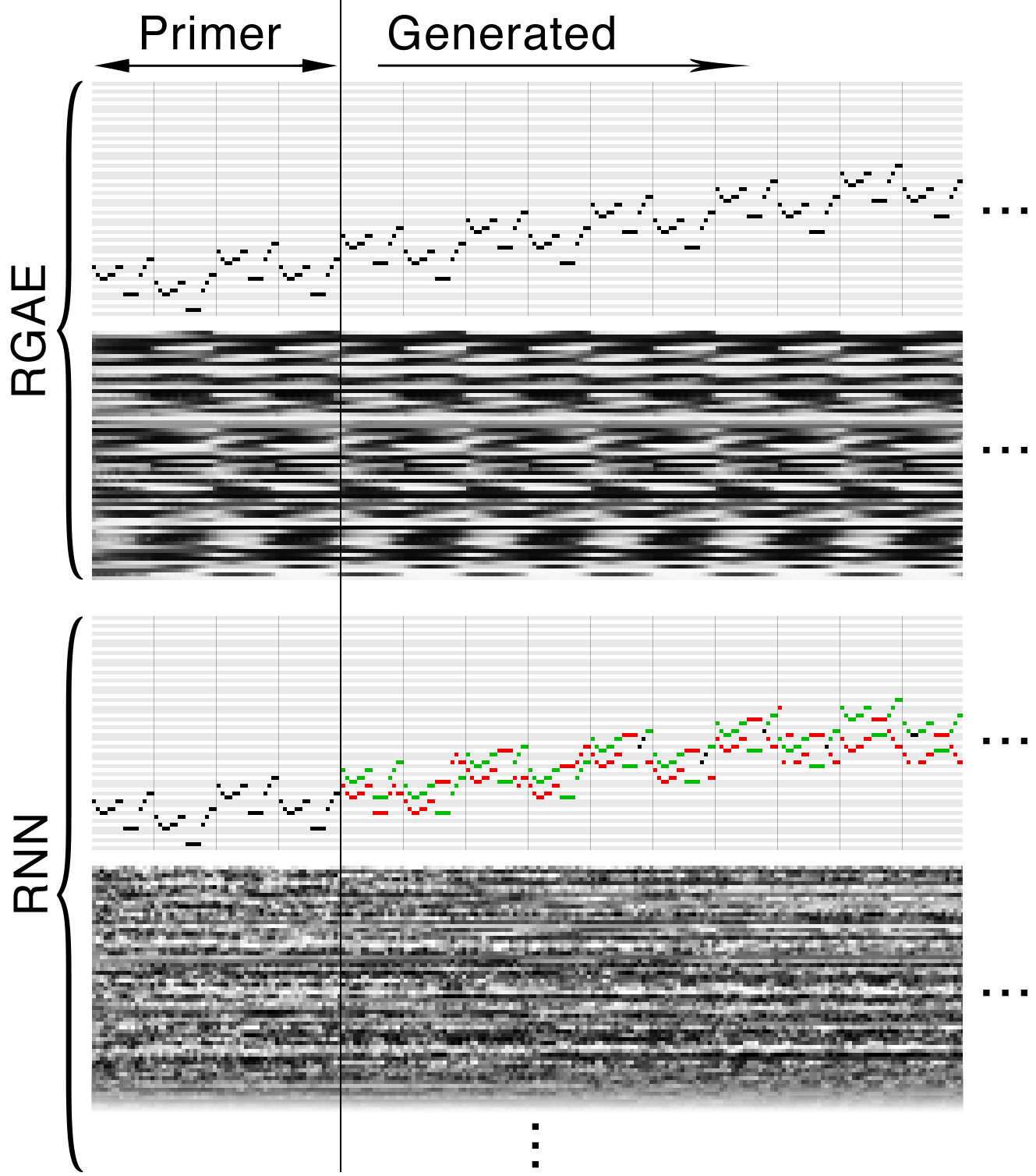} \\
\caption{Generated structure schemes and hidden unit activations of the RGAE and the RNN models after input of a primer indicating the $\{-4,+8,-4,+8,\dots \}$ scheme, realized with melodies of length 16 not contained in the train set. Black notes indicate correct continuation, green notes indicate false negatives, red notes indicate false positives. Hidden units activations of the RNN are pruned due to space limitation.}
\label{fig:seq-continued}
\vspace{-.7cm}
\end{center}
\end{figure}

Table \ref{tab:quant-res-struct} shows the quantitative results of the experiment, and Figure \ref{fig:prec-boxpl} shows a box plot comparing the precisions of the two models.
With an average precision of $99.43\%$ percent, where $92\%$ of all examples are flawlessly continued, the RGAE shows remarkable stability in continuing the structure scheme realizations.
The cross-entropy of the RGAE is about two orders of magnitude lower than that of the RNN.
In Figure \ref{fig:seq-continued}, a specific example of this sequence continuation task is depicted.
Note that the hidden unit activations of the RGAE are more regular because they only represent copy-and-shift operations instead of the musical texture itself (as it is the case for the RNN).
The most challenging part for the RGAE is counting, in order to change the copy operation (i.e., transposition distance) at the right time (in fact, at most of the incorrectly continued sequences, the RGAE miscounted by one time step).
It is important to note that the hidden unit activations of the RNN portion are identical for identical schemes, because they operate on transformations between events, rather than on the events themselves (i.e., they are largely content-invariant).

\begin{table}
\centering
\begin{tabular}{lllll}
\toprule
Model & Pr (\%) & $> 99\%$ & CE & \# Params \\
\midrule
RNN & $41.38$ & $6.67$ & $10.10$ & $\sim 2\,300\,000$ \\
RGAE & $\mathbf{99.43}$ & $\mathbf{92.00}$ & $\mathbf{0.16}$ & $\sim \mathbf{600\,000}$ \\
\bottomrule
\end{tabular}
\caption{Results of the structure learning task. Average precision (Pr), percentage of continuations above $99\%$ precision, cross-entropy (CE) and number of parameters of the respective model.}
\label{tab:quant-res-struct}
\vspace{-.3cm}
\end{table}


\section{Conclusion and future work}\label{sec:conclusion}

The principle of modeling sequences of first-order derivatives in music is a compelling concept with the potential to solve two persistent problems in MIR: Learning trans-position-invariant interval representations, and learning representations of (chromatically transposed) repetition structure.
The proposed model is conceptually simple and can be trained as a generative model in sequence learning tasks.

Moreover, the RGAE can act as a building block for more complex architectures, in order to extend its capabilities.
For example, the temporal lookback window could be greatly extended by employing the RGAE on top of a (dilated) convolutional network, enabling it to learn higher-level repetition structure.
In another variant, an RGAE could be employed on top of an RNN.
Applied to music, the RNN would provide the RGAE with representations of important, absolute reference pitches (e.g., the tonic of a scale, or the root note of a chord), and the RGAE could learn sequences of intervals in relation to them.
Another interesting architecture would involve stacking more than one RGAE on top of one another to learn higher-order derivatives, for example, variations between mutually transposed parts in music.

The RGAE, however, is not limited to the symbolic, monophonic, domain of music.
We show in \cite{lattner2018learning} that a GAE can also operate in the spectral domain of audio and in polyphonic symbolic music.
Finally, we note that the RGAE is general enough to be applicable to other domains where the derivatives of functions are of higher importance than their absolute course.
Possible applications include modeling temporal progressions of changes in loudness, tempo, mood, information density curves, and other musical properties, modeling moving or rotating objects, camera movements in video recordings, and signals in the time domain.

\section{Acknowledgments}
This research was supported by the EU FP7 (project \linebreak Lrn2Cre8, FET grant number 610859), and the European Research Council (project CON ESPRESSIONE, ERC grant number 670035). We thank Srikanth Cherla for providing us with the source code of the RTDRBM model \cite{cherla2016neural}.

\makeatother

\bibliography{bib/bib_mg,bib/bib_cc,bib/bib_sl}

\begin{thebibliography}{10}

\bibitem{cherla2016neural}
Srikanth Cherla.
\newblock {\em Neural Probabilistic Models for Melody Prediction, Sequence
  Labelling and Classification}.
\newblock PhD thesis, City, University of London, 2016.

\bibitem{cho2014properties}
Kyunghyun Cho, Bart van Merri{\"e}nboer, Dzmitry Bahdanau, and Yoshua Bengio.
\newblock On the properties of neural machine translation: Encoder--decoder
  approaches.
\newblock {\em Syntax, Semantics and Structure in Statistical Translation},
  page 103, 2014.

\bibitem{chung2014empirical}
Junyoung Chung, Caglar Gulcehre, Kyunghyun Cho, and Yoshua Bengio.
\newblock Empirical evaluation of gated recurrent neural networks on sequence
  modeling.
\newblock {\em arXiv preprint arXiv:1412.3555}, 2014.

\bibitem{collins2016developing}
Tom Collins, Robin~C. Laney, Alistair Willis, and Paul~H. Garthwaite.
\newblock Developing and evaluating computational models of musical style.
\newblock {\em Artificial Intelligence for Engineering Design, Analysis and
  Manufacturing}, 30(1):16--43, 2016.

\bibitem{conklin_semiotic}
Darrell Conklin.
\newblock Chord sequence generation with semiotic patterns.
\newblock {\em Journal of Mathematics and Music}, 10(2):92--106, 2016.

\bibitem{eigenfeldt2013evolving}
Arne Eigenfeldt and Philippe Pasquier.
\newblock Evolving structures for electronic dance music.
\newblock In {\em Genetic and Evolutionary Computation Conference, {GECCO} '13,
  {Amsterdam, The Netherlands, July} 6-10, 2013}, pages 319--326. ACM, 2013.

\bibitem{herremans2016morpheus}
Dorien Herremans and Elaine Chew.
\newblock Morpheu{S}: {A}utomatic music generation with recurrent pattern
  constraints and tension profiles.
\newblock In {\em Proceedings of the {IEEE} Region 10 Conference ({TENCON}),
  {Singapore, November 22-25}, 2016}, pages 282--285. IEEE, 2016.

\bibitem{hinton2012neural}
Geoffrey Hinton, Nitish Srivastava, and Kevin Swersky.
\newblock Neural networks for machine learning lecture 6a overview of
  mini-batch gradient descent, 2012.

\bibitem{hochreiter1997long}
Sepp Hochreiter and J{\"{u}}rgen Schmidhuber.
\newblock Long short-term memory.
\newblock {\em Neural Computation}, 9(8):1735--1780, 1997.

\bibitem{DBLP:conf/ismir/LanghabelLTR17}
Jonas Langhabel, Robert Lieck, Marc Toussaint, and Martin Rohrmeier.
\newblock Feature discovery for sequential prediction of monophonic music.
\newblock In Sally~Jo Cunningham, Zhiyao Duan, Xiao Hu, and Douglas Turnbull,
  editors, {\em Proceedings of the 18th International Society for Music
  Information Retrieval Conference, {ISMIR} 2017, Suzhou, China, October 23-27,
  2017}, pages 649--656, 2017.

\bibitem{lattner2017relations}
Stefan Lattner and Maarten Grachten.
\newblock Learning transformations of musical material using gated
  autoencoders.
\newblock In {\em Proceedings of the 2nd Conference on Computer Simulation of
  Musical Creativity, {CSMC 2017, Milton Keynes, UK, September} 11-13, 2017},
  2017.

\bibitem{lattnergeneration}
Stefan Lattner, Maarten Grachten, and Gerhard Widmer.
\newblock Imposing higher-level structure in polyphonic music generation using
  convolutional restricted {B}oltzmann machines and constraints.
\newblock {\em Journal of Creative Music Systems}, 3(1), 2018.

\bibitem{lattner2018learning}
Stefan Lattner, Maarten Grachten, and Gerhard Widmer.
\newblock Learning transposition-invariant interval features from symbolic
  music and audio.
\newblock In {\em Proceedings of the 19th International Society for Music
  Information Retrieval Conference, {ISMIR} 2018, {Paris, France, September}
  23-27}, 2018.

\bibitem{Lee:2007uz}
Honglak Lee, Chaitanya Ekanadham, and Andrew~Y. Ng.
\newblock Sparse deep belief net model for visual area {V2}.
\newblock In John~C. Platt, Daphne Koller, Yoram Singer, and Sam~T. Roweis,
  editors, {\em Proceedings of the Twenty-First Annual Conference on Neural
  Information Processing Systems, {Vancouver, British Columbia, Canada,
  December} 3-6, 2007}, pages 873--880. Curran Associates, Inc., 2007.

\bibitem{memisevic2011gradient}
Roland Memisevic.
\newblock Gradient-based learning of higher-order image features.
\newblock In {\em {IEEE} International Conference on Computer Vision ({ICCV}),
  2011}, pages 1591--1598. IEEE, 2011.

\bibitem{memisevic2013aperture}
Roland Memisevic and Georgios Exarchakis.
\newblock Learning invariant features by harnessing the aperture problem.
\newblock In {\em ICML (3)}, pages 100--108, 2013.

\bibitem{memisevic2007unsupervised}
Roland Memisevic and Geoffrey Hinton.
\newblock Unsupervised learning of image transformations.
\newblock In {\em {IEEE} Conference on Computer Vision and Pattern Recognition,
  2007. {CVPR}.}, pages 1--8. IEEE, 2007.

\bibitem{memisevic2010learning}
Roland Memisevic and Geoffrey~E Hinton.
\newblock Learning to represent spatial transformations with factored
  higher-order {B}oltzmann machines.
\newblock {\em Neural Computation}, 22(6):1473--1492, 2010.

\bibitem{michalski2014modeling}
Vincent Michalski, Roland Memisevic, and Kishore Konda.
\newblock "modeling deep temporal dependencies with recurrent grammar cells".
\newblock In {\em Advances in neural information processing systems}, pages
  1925--1933, 2014.

\bibitem{pachet2017sampling}
Fran{\c{c}}ois Pachet, Sony~CSL Paris, Alexandre Papadopoulos, and Pierre Roy.
\newblock Sampling variations of sequences for structured music generation.
\newblock In {\em Proceedings of the 18th International Society for Music
  Information Retrieval Conference}, pages 167--173, 2017.

\bibitem{pearce2004methods}
Marcus Pearce, Darrell Conklin, and Geraint Wiggins.
\newblock Methods for combining statistical models of music.
\newblock In {\em International Symposium on Computer Music Modeling and
  Retrieval}, pages 295--312. Springer, 2004.

\bibitem{pearce2004improved}
Marcus Pearce and Geraint Wiggins.
\newblock Improved methods for statistical modelling of monophonic music.
\newblock {\em Journal of New Music Research}, 33(4):367--385, 2004.

\bibitem{pearce2005construction}
Marcus~Thomas Pearce.
\newblock {\em The construction and evaluation of statistical models of melodic
  structure in music perception and composition}.
\newblock PhD thesis, City University London, 2005.

\bibitem{TheEssenFolksongC:1995um}
Helmut Schaffrath.
\newblock {The Essen Folksong Collection in Kern Format.}
\newblock In David Huron, editor, {\em Database containing , folksong
  transcriptions in the Kern format and a -page research guide computer
  database}. Menlo Park, CA, 1995.

\bibitem{schluter2011music}
Jan Schlueter and Christian Osendorfer.
\newblock Music similarity estimation with the mean-covariance restricted
  {B}oltzmann machine.
\newblock In {\em 10th International Conference on Machine Learning and
  Applications and Workshops ({ICMLA}), 2011}, volume~2, pages 118--123. IEEE,
  2011.

\bibitem{shannon2001mathematical}
Claude~Elwood Shannon.
\newblock A mathematical theory of communication.
\newblock {\em Bell System Technical Journal}, 27:379--423, 623--656, July
  1948.

\bibitem{DBLP:conf/nips/SutskeverHT08}
Ilya Sutskever, Geoffrey~E. Hinton, and Graham~W. Taylor.
\newblock The recurrent temporal restricted {B}oltzmann machine.
\newblock In Daphne Koller, Dale Schuurmans, Yoshua Bengio, and L{\'{e}}on
  Bottou, editors, {\em Advances in Neural Information Processing Systems 21,
  Proceedings of the Twenty-Second Annual Conference on Neural Information
  Processing Systems, Vancouver, British Columbia, Canada, December 8-11,
  2008}, pages 1601--1608. Curran Associates, Inc., 2008.

\bibitem{vincent2010stacked}
Pascal Vincent, Hugo Larochelle, Isabelle Lajoie, Yoshua Bengio, and
  Pierre-Antoine Manzagol.
\newblock Stacked denoising autoencoders: Learning useful representations in a
  deep network with a local denoising criterion.
\newblock {\em Journal of Machine Learning Research}, 11(Dec):3371--3408, 2010.

\bibitem{widmer2003discovering}
Gerhard Widmer.
\newblock Discovering simple rules in complex data: A meta-learning algorithm
  and some surprising musical discoveries.
\newblock {\em Artificial Intelligence}, 146(2):129--148, 2003.

\end{thebibliography}

\end{document}